\documentclass[sigconf]{acmart}

\usepackage{booktabs} 

\usepackage{siunitx}
\usepackage{multirow}
\usepackage{adjustbox}
\usepackage{color}
\usepackage[utf8]{inputenc}
\usepackage[T1]{fontenc}
\usepackage{balance}

\DeclareMathOperator*{\argmin}{arg\,min}

\theoremstyle{definition}
\newtheorem{definition}{Definition}

\copyrightyear{2018}
\acmYear{2018} 
\setcopyright{acmlicensed}
\acmConference[MM '18]{2018 ACM Multimedia Conference}{October 22--26, 2018}{Seoul, Republic of Korea}
\acmBooktitle{2018 ACM Multimedia Conference (MM '18), October 22--26, 2018, Seoul, Republic of Korea}
\acmPrice{15.00}
\acmDOI{10.1145/3240508.3240665}
\acmISBN{978-1-4503-5665-7/18/10}

\begin{document}
\title{Temporal Cross-Media Retrieval with Soft-Smoothing~\textsuperscript{*}}

\thanks{* Please cite the ACM MM 2018 version of this paper.}

\author{David Semedo}
\affiliation{%
\institution{NOVA LINCS}
\institution{Universidade NOVA de Lisboa}
\streetaddress{}
\city{Caparica}
\country{Portugal}
}
\email{df.semedo@campus.fct.unl.pt}

\author{Joao Magalhaes}
\affiliation{%
\institution{NOVA LINCS}
\institution{Universidade NOVA de Lisboa}
\streetaddress{}
\city{Caparica}
\country{Portugal}}
\email{jm.magalhaes@fct.unl.pt}

\renewcommand{\shortauthors}{D. Semedo et al.}

\begin{abstract}
Multimedia information have strong temporal correlations that shape the way modalities co-occur over time. 
In this paper we study the dynamic nature of multimedia and social-media information, where the temporal dimension emerges as a strong source of evidence for learning the \textit{temporal correlations across visual and textual modalities}. 
So far, cross-media retrieval models, explored the correlations between different modalities (e.g. text and image) to learn a common subspace, in which semantically similar instances lie in the same neighbourhood.
Building on such knowledge, we propose a novel temporal cross-media neural architecture, that departs from standard cross-media methods, by explicitly accounting for the temporal dimension through temporal subspace learning. 
The model is \textit{softly-constrained with temporal and inter-modality constraints} that guide the new subspace learning task by favouring temporal correlations between semantically similar and temporally close instances. 
Experiments on three distinct datasets show that accounting for time turns out to be important for cross-media retrieval. Namely, the proposed method outperforms a set of baselines on the task of temporal cross-media retrieval, demonstrating its effectiveness for performing temporal subspace learning.
\end{abstract}

\ccsdesc[500]{Information systems~Multimedia and multimodal retrieval}

    \keywords{Cross-media; temporal cross-media; temporal smoothing; multimedia retrieval}

\maketitle

\graphicspath{{figs/}}

\section{Introduction}
\begin{figure}[t]
    \centering
    \includegraphics[width=\linewidth]{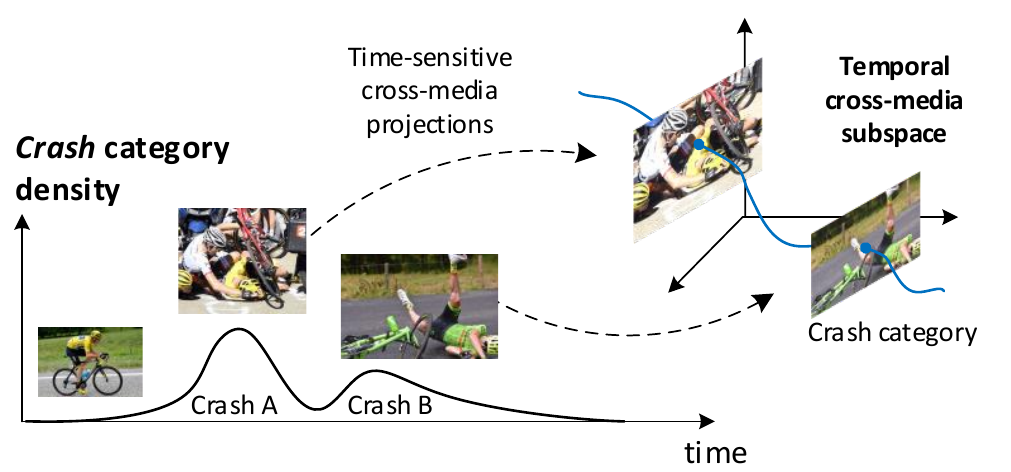}
\caption{Temporal Dynamics of semantic category \emph{Crash} (TDF2016), and temporal pairwise variations with corresponding visual elements.}
  \label{fig:temporal_overview}
  \vspace{-5mm}
\end{figure}

Solid advances have been proposed to annotate media with complex keywords \cite{DBLP:journals/corr/SimonyanZ14a} and retrieve information across different modalities \cite{Rasiwasia:2010:NAC:1873951.1873987, Wang:2017:ACR:3123266.3123326, 7780910}.
Cross-media models enable the retrieval of content from a target modality (e.g. text) given another modality (e.g. image). 
The field has been extensively researched and attracted many contributions~\cite{Rasiwasia:2010:NAC:1873951.1873987,Feng:2014:CRC:2647868.2654902, 7298966}, with the most widely used approach being common space learning~\cite{7930456}. 
Among the pioneering works, is the work of~\citet{Rasiwasia:2010:NAC:1873951.1873987} which leveraged on the Canonical Correlation Analysis~\cite{hotelling36cca} (CCA) algorithm to learn a multimodal linearly correlated subspace. The field then evolved to the adoption of neural-network methods, which allow the learning of complex non-linear projections~\cite{Feng:2014:CRC:2647868.2654902, 7298966, Peng:2016:CSR:3061053.3061157}, through the composition of several non-linear functions. Recently,~\citet{Wang:2017:ACR:3123266.3123326} elegantly formulated the cross-media subspace learning problem under an adversarial learning framework. 
A common assumption of previous works is the static corpora assumption, thus, overlooking temporal correlations between \emph{visual-textual} pairs. 
Looking at Figure~\ref{fig:temporal_overview} one can see that images and texts are not static across the corpus timespan, with the later being reflected on \emph{visual-textual} pairs in the form of temporal correlations. 
These lead to the existence of cross-modal pairwise correlations that change over time.

Modelling such dynamic cross-media relations over time raises many challenges for current state-of-the-art cross-media analysis. 
However, it is possible to see how numerous works have leveraged on the dynamics of social media for diverse tasks, such as Emerging Events Tracking~\cite{Lau2012OnlineTA}, Dynamic Sentiment Analysis~\cite{Tsytsarau:2014:DNE:2623330.2623670}, Natural Disasters Prediction~\cite{Sakaki:2010:EST:1772690.1772777}, among others. 
In all these works, content temporal relevance insights proved to be crucial. Namely, they exploit the fact that user contributed content, from certain topics, follows some temporal pattern. 
For instance, as illustrated in figure~\ref{fig:temporal_overview}, the semantic category \emph{Crash}, has two modes along the temporal dimension with different cross-modal correlations, corresponding to a multimodal variation of the category \emph{crash} across the corpus timespan.

In this paper, we \textit{hypothesise that, for dynamic corpora, both visual and textual modalities change over time}, creating the need for time-sensitive cross-media retrieval. 
To assess this hypothesis, new time-sensitive cross-media subspace learning approaches will be investigated.
The goal is to learn effective projections where cross-media patterns are decomposed over the temporal dimension, thus, proving the hypothesis.
We take a step further by seeking a space where semantically similar and \emph{mutually temporally related instances}, lie in the same neighbourhood.

We propose \emph{TempXNet}, a novel Temporal Cross-Media Neural Architecture, to learn projections for both textual and visual data, while modelling temporal correlations between modalities. 
A temporal subspace learning approach is devised, by enforcing temporal constraints between semantically similar instances, in order to learn \textit{time-sensitive modality projections}. This novel strategy will enable effective retrieval in a temporally-aware cross-media subspace. 
Many cross-media works have already incorporated external information (e.g. semantic categories information), when performing subspace learning, breaking instances' pairwise coupling, in order to better capture \emph{visual-textual} correlations~\cite{Gong:2014:MES:2584252.2584265,8013822,8019310}. Accordingly, in the proposed method the underlying dynamics of instances are captured by considering two distinct temporal correlation modelling approaches. 
These model the intra-category temporal correlations at two levels of granularity: at the document's timestamp level and at individual word's level.
The key aspects of TempXNet are summarised as follows:
\begin{enumerate}
    \item The proposed model is flexible enough to support cross-media temporal correlations following parametric, non- parametric and latent-variable distributions.
    \item The formulation of the subspace learning objective function, that captures the temporal correlation, is a differentiable function which can be conveniently optimised by gradient-based algorithms.
\end{enumerate}
Experiments on three datasets illustrate the importance of considering temporal correlations in cross-media retrieval tasks. TempXNet outperformed recent cross-media retrieval baselines. We further evaluated the behaviour of different temporal distributions to model the cross-media dynamics of social media content.

\section{Related Work}
The literature on modelling and incorporating temporal aspects for multimodal retrieval is very scarce. A pioneer approach was devised by~\citet{Kim:2013:TWI:2433396.2433417}, where temporal clues are used to improve search relevance at query time, modelling content streams using a multi-task regression on multivariate point processes. \emph{Visual-textual} pairs are treated as random events in time and space. Image ranking, under this framework is improved by using a multi-task learning approach that considers multiple image descriptors when capturing temporal correlations.~\citet{10.1007/978-3-642-41181-6_73} evaluated the value of temporal information, such as tag frequency, for the task of image annotation and retrieval. The authors confirm that some tags reveal a dynamic behaviour that was found to be aligned with Google search trends, thus supporting our hypothesis regarding the behaviour of \emph{visual-textual} pairs, on (social) dynamic corpus. On the other hand, for orthogonal tasks but directly dealing with social corpora, time, or more specifically, temporal relevance of elements, has been exploited~\cite{Lau2012OnlineTA,Tsytsarau:2014:DNE:2623330.2623670,Sakaki:2010:EST:1772690.1772777}. Our hypothesis is directly inspired and supported by the findings of such works, which successfully exploited temporal insights, encoded on social corpora.

For static collections, the task of cross-media retrieval, between visual and textual modalities, has been extensively researched~\cite{Rasiwasia:2010:NAC:1873951.1873987,Ngiam:2011:MDL:3104482.3104569,Srivastava12learningrepresentations, Feng:2014:CRC:2647868.2654902, 7298966,Fan:2017:CRL:3123266.3123369,8013822,Wang:2017:ACR:3123266.3123326}. Namely, neural network methods have proved to be highly effective at learning non-linear projections. 
~\citet{Feng:2014:CRC:2647868.2654902} proposed the Correlation Autoencoder (Corr-AE), which is comprised by two Autoencoders (one for each modality), whose intermediate layers are tied by a similarity measure. The loss function is then defined by the autoencoder reconstruction error and an additional cost term, measuring the correlation error.~\citet {7298966} leveraged on Deep Canonical Correlation Analysis (DCCA) to match images and text. DCCA exploits the fact that the CCA objective function can be formulated based on a  matrix trace norm.
~\citet{Fan:2017:CRL:3123266.3123369}, combine image global (CNN features) and descriptive (e.g. caption) representations using a network fusion approach to obtain a richer semantic embedding space. Very recently,~\citet{Wang:2017:ACR:3123266.3123326} proposed to learn a common space using an adversarial learning approach.
Inspired by state-of-the-art subspace learning works~\cite{Wang:2017:ACR:3123266.3123326,7298966, Feng:2014:CRC:2647868.2654902, 7298966} we consider neural networks, which have proved to be very effective. Namely, we adopt the two-network (one for each modality) base architecture for projection learning, predominant across multiple state-of-the-art works. 

Joint representation learning for video sequences, with well aligned visual and audio modalities, has been actively researched~\cite{7780758, DBLP:conf/cvpr/YangRCMBL17, Pan:2016:LDI:3061053.3061155, pmlr-v37-srivastava15}. These are commonly based on temporal methods, such as Recurrent Neural Networks, that are able to capture dependencies on sequences, over time steps. However, these sequences are fundamentally different from temporal correlations of \emph{visual-textual} pairs on social media, as they are assumed to show coherence over time and be perfectly aligned. 

Hence, there is considerable literature~\cite{Lau2012OnlineTA,Tsytsarau:2014:DNE:2623330.2623670,Sakaki:2010:EST:1772690.1772777} outside cross-media retrieval to support this paper hypothesis. Therefore, we explicitly address the problem and propose a temporal cross-media subspace learning approach for dynamic corpora, in which latent temporal correlations are intrinsically accounted.

\section{Temporal Cross-Media Subspace}
The main hypothesis we wish to investigate (and quantify) is that both visual and textual modality correlations change over time. This is supported by the existence of dynamic visual-textual pairs (Figure~\ref{fig:temporal_overview}) originating multi-dimensional correlations among the temporal, visual and textual dimensions of the problem's data. Consequently, we argue that temporal correlations between instances of a same semantic category should lead to the investigation of new subspaces that capture such data interactions.

\begin{figure*}
\centering
  \includegraphics[width=0.9\linewidth]{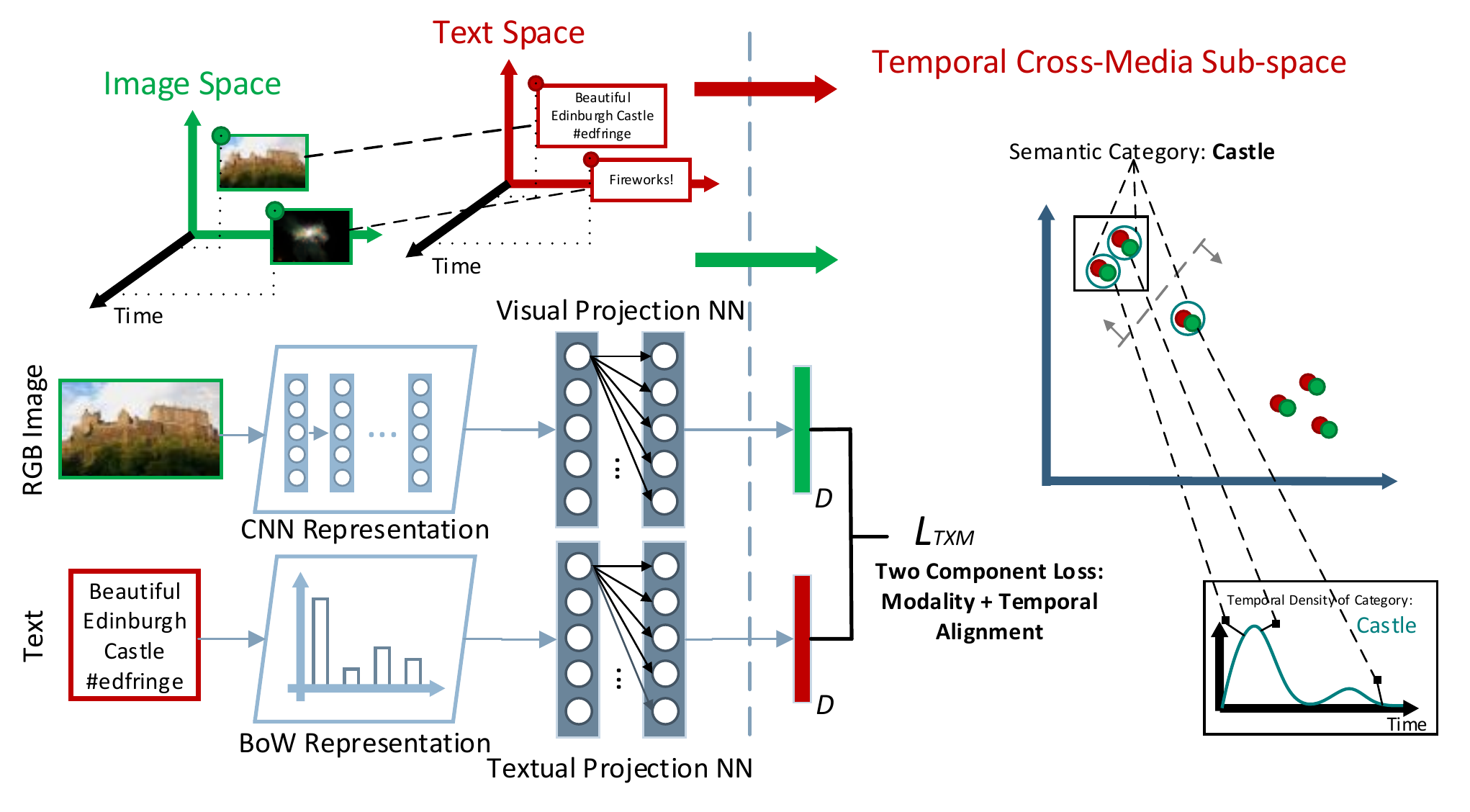}
  \caption{Temporal Cross-Media subspace learning overview. Visual (green) and textual (red) instances are mapped to a $D$ dimensional cross-media space. The space is perturbed to approximate temporally correlated instances, of a same category, and to separate uncorrelated ones. Best viewed in color.}
	\label{fig:framework_overview}
	
\end{figure*}

\subsection{Subspace Definition}
Let $\mathcal{C}$ be a corpus of timestamped documents, with a timespan $TS = [t_s, t_f]$, where $t_s$ and $t_f$ are the first and last  span instants respectively, where each instance $d^i \in \mathcal{C}$ is defined as 
\begin{equation}
d^i=( x_I^i, x_T^i, t^i),
\end{equation}
where $x_I^i \in \mathcal{R}^{D_I}$ and $x_T^i \in \mathcal{R}^{D_T}$ are the instance's image $d_I$ and text $d_T$ feature representations, respectively, and  $t^i$ the instance's timestamp. Accordingly, $D_I$ and $D_T$ correspond to the image and text features dimensionality, respectively. We define $V$ as the vocabulary of the text data and $L$ as the set of semantic categories of the corpus $\mathcal{C}$.
Each instance $d^i$ is associated with a set of semantic categories $l^i = \{l_1^i, \ldots, l_j^i, \ldots, l_{|l^i|}^i\}$, such that each $l_j^i \in L$. It follows that each instance $d^i$ can be associated with one or more categories. 
This allows us to introduce the formal definition of temporal cross-media subspace:

\begin{definition}{A \emph{temporal cross-media subspace}} 
refers to a subspace that is learned from timestamped multimedia data to organize data according to their semantic category and temporal correlations across different modalities.
\end{definition}

Formally, the temporal cross-media subspace will have the following properties:
\begin{description}
    \item[Property 1.] Two elements will be maximally correlated in the new subspace if they share at least one semantic category and if they are strongly correlated in time;
    \item[Property 2.] Considering the same semantic category, temporally correlated instances will lie in the same neighbourhood, while temporally uncorrelated instances are expected to lie far apart;
    \item[Property 3.] Complex temporal behaviours will be captured by a function $sim_{temp}$, grounded on a temporal distribution $\theta_{temp}$, which can follow a parametric, non-parametric or latent-variable model.
\end{description}

\subsection{Time-sensitive Cross-Media Projections}
Given the aforementioned definitions, it follows that both $x_I$ and $x_T$ original spaces are dissimilar and obtained without accounting for time. Namely each space may have different dimensionality, semantics and distributions, making them incompatible.
This leads us to the projections:
\begin{equation}
\mathcal{P}_I(\cdot;\theta_I): \mathbb{R}^{D_I} \mapsto \mathbb{R}^D \hspace{1cm} \mathcal{P}_T(\cdot;\theta_T): \mathbb{R}^{D_T} \mapsto \mathbb{R}^D
\end{equation}
mapping images $d_I$ and texts $d_T$ to a common temporal cross-media subspace, with dimensionality $D$, according to the to image and text projections models $\theta_I$ and $\theta_T$, correspond, respectively.

To learn the time-sensitive cross-media projections $\mathcal{P}_{I}(\cdot;\theta_I)$ and $\mathcal{P}_{T}(\cdot;\theta_T)$, it is essential to maximise the correlation in the new subspace between the two modalities, both at the semantic and temporal dimensions. Thus, the projections into the temporal cross-media subspace need to capture the temporal traits of semantic categories, which are grounded on temporal correlations across visual and textual modalities. In practice, we argue for projections that are learned with novel temporally constrained objective functions of the form \begin{equation}
\begin{gathered}
\argmin_{\theta_I, \theta_T} \mathcal{L}_{TXM}(\theta_I, \theta_T) \\
\text{s.t. } \ \  \mathcal{L}_{temp}(\theta_I, \theta_T;\theta_{temp}) = 0,
\label{eq:general_loss}
\end{gathered}
\end{equation}
where $\mathcal{L}_{TXM}$ corresponds to a cross-media loss that minimises the distance over semantically similar representations and maximises the distance between semantically dissimilar instance's representations. The cross-media loss is subject to temporal smoothing constraints, imposed by a temporal factor $\mathcal{L}_{temp}$, grounded on a temporal model $\theta_{temp}$, that enforces the aforementioned properties. As will be detailed later, temporal constraints $\mathcal{L}_{temp}$ are relaxed as an additive smoothing term, added to $\mathcal{L}_{TXM}$.

\subsection{Temporal Cross-Media Retrieval}
Many tasks can be solved in the temporal cross-media subspace, which are not supported by current cross-media models. Given a cross-media query $q$, defined as
\begin{equation}
q=( \langle x_I \rangle,\langle x_T \rangle),
\end{equation}
where one or both modalities may be specified, the goal is to retrieve the set of temporally and semantically correlated instances (from the remaining modality when only one is specified). Our model intrisically encodes temporal correlations on a cross-media space such that it does not require a timestamp as input to project modalities close to temporally correlated instances. Also, it allows marginalising the model along of the visual and textual variables, and obtain temporally grouped data.

In the following sections we will detail how the TempXNet architecture is materialised. Namely the loss function from equation~\ref{eq:general_loss}, the temporal factor $\mathcal{L}_{temp}$ and the neural cross-media architecture.

\section{Temporal Cross-Media Learning}
The goal of temporal cross-media subspace learning is to create a new subspace where semantic and temporal latent correlations, for instances of the same category, are represented at essentially two granularity levels:  1) inter-modality pairwise correlation, and 2) inter-instance correlation.

On the obtained temporal subspace, the encoding of the temporal dimension is achieved by smoothing the aligned representations subspace with a set of temporal constraints imposed on the loss function from equation~\ref{eq:general_loss}. 
The temporal factor term $\mathcal{L}_{temp}$ is backed up by a temporal model $\theta_{temp}$, estimated from the corpus, that provides two temporal insights: 1) instance's \emph{temporal signatures} over the corpus $\mathcal{C}$ timespan, and 2) a smoothed temporal correlation functions, based on the aforementioned temporal signatures. It is important to see how this cost function leads to cross-media projections fundamentally different from previous works, where images and text are grouped in a temporally agnostic manner.

\subsection{Projection Learning}
Apart from the temporal insights, it is crucial to learn effective modality projections, that map original modality vectors to a new space where pairwise (visual and textual modalities) and instance's semantic correlations are represented. Inspired by state-of-the-art cross-media retrieval approaches~\cite{7410369,7780910,Wang:2017:ACR:3123266.3123326}, we formulate $\mathcal{L}_{TXM}$ as a pairwise ranking-loss~\cite{herbrich2000large}, as it has been shown that minimisation of this loss is directly related with the maximisation of nDCG and MAP~\cite{NIPS2009_3708}. Thus,  $\mathcal{L}_{TXM}$ is defined as follows:
\begin{equation}
\begin{split}
     \mathcal{L}&_{TXM}(\theta_I, \theta_T) =\\ 
     &\sum_{i,n} max(0, m-\mathcal{P}_I(x_I^i)\cdot\mathcal{P}_T(x_T^i) + \mathcal{P}_I(x_I^i)\cdot\mathcal{P}_T(x_T^n)) \ \ + \\
     &\sum_{i,n}  max(0, m-\mathcal{P}_T(x_T^i)\cdot\mathcal{P}_I(x_I^i) + \mathcal{P}_T(x_T^i)\cdot\mathcal{P}_I(x_I^n)),
    \label{eq:pairwise_loss}
\end{split}
\end{equation}
where $x_I^n$ and $x_T^n$ are images and texts representations from negative instances, w.r.t. an instance $d^i$. Similarity  between projections is computed by a dot product over two unit-norm, $\ell_2$ normalised vectors, making it equivalent to cosine similarity.

We devise a neural network architecture to perform subspace learning and learn projections  $\mathcal{P}_I(\cdot)$ and\ $\mathcal{P}_T(\cdot)$. Figure~\ref{fig:framework_overview} depicts the neural architecture. Following~\cite{Ngiam:2011:MDL:3104482.3104569,7298966,Fan:2017:CRL:3123266.3123369}, we consider two neural networks to learn non-linear mappings, with $\theta_I$ and $\theta_T$, denoting each sub-network's learnable parameters for image and textual modalities, respectively. Through the composition of several non-linearities, neural networks are able to model complex latent correlations. Thus, for each modality, a feedforward network, comprising 2 fully connected layers is used. The first layer has 1024 dimensions and the second one has $D$ dimensions. For semantically rich image feature representation, a convolutional neural network is prefixed to the input of the visual modality projection network.

Each modality network takes as input the corresponding modality of an instance $d^i$. Namely a visual projection sub-network takes as input the RGB image $d_I^i$, and a textual projection sub-network a bag-of-words representation of the text $d_T^i$. Both original modality representations are embedded onto a new $D$-dimensional subspace.
Both sub-networks are then jointly optimised by minimising the loss function $\mathcal{L}(\theta_I,\theta_T)$, from eq.~\ref{eq:general_loss}. Apart from the training phase, both sub-networks are decoupled and thus can be used independently to map a single modality. 

\subsection{Temporal Cross-Media Soft-Constraints}

Temporal subspace learning properties are enforced over semantically similar instances only, through a set of soft-constraints. Thus, the temporal factor $\mathcal{L}_{temp}$ is defined as:
\begin{equation}
     \mathcal{L}_{temp}(\theta_I, \theta_T) = \lambda\sum_i \mathcal{L}_{temp}(d^i;\theta_I, \theta_T),
    \label{eq:temp_loss}
\end{equation}
where $\lambda$ is an hyper-parameter used to control the influence of the temporal factor. $\mathcal{L}_{temp}(\theta_I, \theta_T)$ is then added to eq.~\ref{eq:general_loss} as a smoothing term. 
The rationale of equation~\ref{eq:temp_loss} is to smooth the model by constraining the learned projections for every instance $d^i$, with temporal soft-constraints. For a single instance $d^i$, let $J = \{j: l^j \cap l^i \neq \emptyset \}$ be the set of positive examples $d^i$ of category of $l^j$. The constraints are:
\begin{itemize}
    \item Temporally correlated instances, with distant cross-modality projections, should have similar projections. Violations to this constraint are captured as follows:
\begin{equation}
\begin{split}
 C1(d^i) = \frac{1}{|J|}\sum_{j \in J}  &sim_{temp}(t^i, t^j;\theta_{temp})\\
 &\cdot (1-sim_{cmod}(d^i, d^j));
\end{split}
\end{equation}
    \item Temporally uncorrelated instances, with close cross-modality projections, should lie far apart, thus having distant projections. Violations to this constraint are captured as follows:
\begin{equation}
\begin{split}
 C2(d^i) = \frac{1}{|J|}\sum_{j \in J} &(1-sim_{temp}(t^i, t^j;\theta_{temp}))\\
 &\cdot sim_{cmod}(d^i, d^j),
\end{split}
\end{equation}
\end{itemize}
where $sim_{temp}(t^i, t^j;\theta_{temp})$, detailed in section~\ref{subsec:temporal_similarities}, is a temporal correlation assessment function that evaluates how correlated in time two instances $d^i$ and $d^j$ are. Finally, $sim_{cmod}(d^i, d^j)$ is a cross-modality similarity function that evaluates how close each modality projection is, w.r.t. to the other modality, on the cross-media subspace. We average pairwise violations to deal with unbalanced positive sets.

The two soft-constraints are then combined as:
\begin{equation}
     \mathcal{L}_{temp}(d^i;\theta_I, \theta_T) =  (C1(d^i) + C2(d^i)).
    \label{eq:temp_loss_instance}
\end{equation}
Essentially, for a given instance $d^i$, $\mathcal{L}_{temp}$ iterates through all the positive instances $d^j$ (sharing at least one semantic category with $d^i$), and computes the two products between temporal and cross-modality distances.

\subsubsection{Cross-modality similarity} Cross-modality similarity\\ $sim_{cmod}$, computed over semantically similar instances of $d^i$, is defined based on the harmonic mean between the cross-modality projections' similarities:
\begin{equation}
     sim_{cmod}(d^i, d^j) = 2\cdot \frac{\mathcal{P}_I(x_I^i)\mathcal{P}_T(x_T^j)\cdot\mathcal{P}_T(x_T^i)\mathcal{P}_I(x_I^j)}{\mathcal{P}_I(x_I^i)\cdot\mathcal{P}_T(x_T^j) + \mathcal{P}_T(x_T^i)\cdot\mathcal{P}_I(x_I^j) +\epsilon}
\label{eq:sim_mod}
\end{equation}
where again, similarity is computed by a dot product $\ell_2$ normalised vectors. A small constant $\epsilon$ is added to the denominator to avoid zero division. Essentially, $sim_{cmod}$ assesses the alignment between the representations obtained by projections $\mathcal{P}_I(\cdot)$ and $\mathcal{P}_T(\cdot)$, over two instances, by equally weighting both modalities' projections.

\subsection{Temporal Soft-Smoothing Functions}
\label{subsec:temporal_similarities}


For semantic categories and words, correlation strength within different instances, is expected to vary over time. We posit that the later is reflected on the dynamic behaviour of each element (a semantic category or a word). On a corpus $\mathcal{C}$, such behaviour is accounted by $\mathcal{L}_{temp}$, through a temporal correlation assessment function $sim_{temp}$. We materialise the later based on two fundamentally different levels: \textit{category} and \textit{word} temporal behaviour.

\subsubsection{Category-based Correlations} 
We propose to assess temporal correlations by directly comparing temporal density distribution $\vec{\phi}_{l}$ of categories. Given $\vec{\phi}_{l}$, we define the temporal density of $l$, at time $t$, as a probability function $p(t|\vec{\phi}_{l})$. Then, Category-based correlations are then defined as:
\begin{equation}
\begin{split}
     sim_{temp}(t^i, t^j)=p(t^i|\vec{\phi}_{l})\cdot p(t^j|\vec{\phi}_{l}),
\label{eq:label_density_estimation}
\end{split}
\end{equation}
such that $p(t|\vec{\phi}_{l})$ corresponds to the relevance of label $l$, at time $t$. When two instances share more than one label, we consider the value of the label that maximises $sim_{temp}$. Kernel Density Estimation (KDE), with a Gaussian Kernel, is used to obtain a smoothed estimation of $\vec{\phi}_{l}$. The bandwidth $h$ hyper-parameter is used to control the smoothness of the estimated density.

\subsubsection{Topic-based Correlations}
Individual word's dynamic behaviour provides a richer insight regarding \emph{visual-textual} temporal pairs correlations. Namely, it is expected that some \emph{domain-specific} words will have a rich dynamic behaviour, depicting temporal correlations, which should be accounted. Such correlations are also much more fine-grained, when compared to individual semantic categories. 

We model temporal density distributions $\vec{\phi}_{w}$ of each word $w\in x^i_T$ of a instance $d$, through a dynamic topic modelling approach. Topic-based correlations are then defined as:
\begin{equation}
\begin{split}
     sim_{temp}(t^i, t^j)= & \ \ \ p(t^j|x^i_T) =\prod_{w\in x^i_T} p(t^j|\vec{\phi}_{w}),
\label{eq:word_density_estimation}
\end{split}
\end{equation}
such that $p(t|\vec{\phi}_{w})$ corresponds to the density of word $w$, at time $t$. Essentially, equation~\ref{eq:word_density_estimation} measures temporal correlation by comparing the temporal density of words in $d_T^i$, at timestamp $t^j$ of document $d^j$. 

To estimate $\vec{\phi}_{w}$, we resort to Dynamic Topic Modelling (DTM), that intrinsically accounts for the time evolution of latent topics. From the DTMs methods family, we consider Dynamic Latent Dirichlet Allocation~\cite{Blei:2006:DTM:1143844.1143859} (D-LDA) to study temporal behaviours of individual words.
The LDA~\cite{Blei:2003:LDA:944919.944937} method represents documents as a finite mixture over a set of estimated latent topics, where each latent topic is characterised by a distribution over words, from which documents are assumed to be generated from. It consists of an \emph{exchangeable} model, as joint probabilities over words are invariant to permutations. D-LDA takes a step further by explicitly addressing topic evolution and dropping the exchangeable property. Documents are arranged into a set of \emph{time slices} and for each time slice, documents are modelled using a $P$-component topic model (LDA), where its latent topics at time slice $t$ evolve from latent topics of slice $t-1$. D-LDA is applied to the corpus $\mathcal{C}$ with time slices referring to individual days. 

For each word and latent-topic $p$, a temporal density curve $\vec{\phi}_{wp}$ is estimated. The element-wise mean over all latent-topics is computed as $\vec{\phi}_{w} = \sum_{p=0}^{P}\vec{\phi}_{wp}$, and normalised. Then, for a given word $w$:
\begin{equation}
p(t|\vec{\phi}_{w}) = f_{dlda}(t,w) = \vec{\phi}_{w}(t),
\end{equation}
where $\vec{\phi}_{w}(t)$ denotes the estimated averaged temporal density, at time instant $t$, across all topics. Given that we average each $\vec{\phi}_{wp}$ over the $P$ latent-topics and that each word $w$ reveals different behaviours on each latent-topic, we obtain a model that captures word variations w.r.t. word correlations with groups of words, over time.

\section{Evaluation}

\subsection{Datasets}

\subsubsection{NUS-WIDE~\cite{nus-wide-civr09}}
Comprised of a total of 269,648 images from the Flickr network, annotated with a total of 81 semantic categories. We crawl images' metadata and stored the \emph{datetaken} field to be used as timestamp. Each image has multiple tags and may belong to multiple semantic categories. We consider the 1000 more frequent tags for text representation. Images that are missing, do not have associated tags, or without timestamp are excluded. We only keep images from year 1999 to 2009\footnotemark, resulting in a total of 169,283 images. We use the NUS-WIDE dataset for temporal cross-media as some tags have been shown to reveal a dynamic behaviour~\cite{10.1007/978-3-642-41181-6_73}. Train, validation and test splits comprise 129,500, 22,854 and 17,112 instances, respectively.
\footnotetext{The dataset was released on 2009, with its distribution having a mean of $2006.69\pm1.175$.}

\subsubsection{SocialStories Dataset}
\footnote{\url{http://datasets.novasearch.org/}}
This dataset consists of a collection of social media documents covering a large number of sub-events about two distinct major events of interest for the general public. In particular, we considered Twitter as a source of social media content.
We specifically considered events that span over multiple days and that contain considerable amounts of diverse visual material. These are expected to have strong temporal correlation across modalities with respect to its semantics. 
Taking the aforementioned aspects into account, we selected the following events:
\begin{description}
	\item[Edinburgh Festival 2016 (EdFest 2016)]~\footnote{\url{https://www.eif.co.uk/}} - Consists of a celebration of the performing arts, gathering dance, opera, music and theatre performers from all over the world. The event takes place in Edinburgh, Scotland and has a duration of 3 weeks in August. The dataset contains 82,348 documents where 1,186 were annotated with 13 semantic categories (\emph{Audience/ Crowd}, \emph{Castle},  \emph{Selfies/Group Photos/Posing}, \emph{Fireworks}, \emph{Music}, \emph{Streets of Edinburgh}, \emph{Food}, \emph{Dance/Dancing}, \emph{Show/Performance},  \emph{Building(s)/Monuments}, \emph{Sky/Clouds}, \emph{Person},  \emph{Water}).
	\item[Le Tour de France 2016  (TDF 2016)]~\footnote{\url{http://www.letour.com/}} - Consists of one of the main road cycling race competitions. The event takes place in France (day 1-8, 11-17, 20-23 ), Spain (day 9), Andorra (day 9-11), Switzerland (day 17-19), and has a duration of 23 days in July. The dataset contains 325,074 documents where 747 were annotated with 13 semantic categories (\emph{Spectators}, \emph{Bicycle/Pedalling}, \emph{Road},  \emph{Yellow-Jersey}, \emph{Car/Truck}, \emph{Peloton},  \emph{Crash}, \emph{Field(s)/ Mountain(s)} \emph{Buildings/Monument(s)}  \emph{Food} \emph{Sky/Clouds}, \emph{Water} and \emph{Person}).
\end{description}

After crawling content with event specific hashtags and seeds, we applied a set of content filtering techniques~\cite{McMinn:2013:BLC:2505515.2505695,benevenuto2010detecting} to discard SPAM and annotated documents event-specific semantic categories. Annotators were asked to annotate media documents (image and text) with one or more categories. An additional \emph{None} category is shown, when none of the categories apply to the instance. 
We obtained a total of 1186 and 747 annotated pairs, with an average of $3.0\pm1.47$ and $2.4\pm1.26$ categories per instance, for EdFest2016 and TDF2016, respectively. 

\subsection{Methodology}
We evaluate the retrieval performance using mean Average Precision (\emph{mAP@K}), which is the standard evaluation metric for cross-media retrieval~\cite{Rasiwasia:2010:NAC:1873951.1873987, Wang2016ACS, Feng:2014:CRC:2647868.2654902,Wang:2017:ACR:3123266.3123326, 7298966} and normalized Discounted Cumulative Gain ($nDCG@K$). We follow~\cite{Feng:2014:CRC:2647868.2654902, Wang:2017:ACR:3123266.3123326} and set $K=50$. For \emph{mAP@K}, an instance is relevant if it shares at least one category. For $nDCG@K$, relevance is defined as the number of common categories. Cross-media retrieval methods are evaluated in two tasks: 1) \emph{Image-to-Text} retrieval (I-T) and 2) \emph{Text-to-Image} (T-I) retrieval. We complement our evaluation with a qualitative analysis.

\subsection{Implementation Details}
Networks are jointly trained using SGD, with $0.9$ momentum, and a learning rate of $\eta = \num{5e-3}$, with a decay of $\num{1e-6}$, and each gradient update step being $\theta_I = \theta_I -\eta\frac{1}{s}\nabla_{\theta_I}(\mathcal{L}_{TXM} + \mathcal{L}_{temp})$ and $\theta_T = \theta_T -\eta\frac{1}{s}\nabla_{\theta_T}(\mathcal{L}_{TXM} + \mathcal{L}_{temp})$.
Early stopping is used to avoid overfitting. Mini-batch size is set to $10,000$, and $64$, for NUS-WIDE and SocialStories, respectively, and the total number of epochs is set to $25$. 
For each neuron, we use \emph{tanh} non-linearities. 
Pre-trained ResNet-50~\cite{He_2016_CVPR}, with the last fully connected layer removed (softmax), is used for image representation. In SocialStories, DLDA was trained on the full un-annotated dataset. The number of latent topics $P$ is set to $10$ through cross-validation. We set $D=100$, $\lambda=1.0$,  pairwise-ranking loss margin $m=1.0$, and KDE bandwidth $h=1$. We adopt the TF-IDF bag-of-words representation for texts. For baselines, image representations are obtained from the same pre-trained ResNet-50 CNN.

\subsection{Experiments}

\vspace{3mm}
\textbf{Cross-Modal Retrieval.} We start by evaluating the proposed subspace learning model, TempXNet, with  each of the  three distinct temporal correlations on the task of cross-media retrieval. Namely, we evaluate semantic category-based temporal correlations,\\~\textbf{TempXNet-Cat} (eq.~\ref{eq:label_density_estimation}); and latent-topic word-based temporal correlations, \textbf{TempXNet-Lat} (eq.~\ref{eq:word_density_estimation}).
We also evaluate an additional straightforward temporal correlation, \textbf{TempXNet-Rec}, in which the correlation corresponds to how close in time two instances are $      sim_{temp}(t^i, t^j)= e^{-|t^i-t^j|/h}$, where $h=0.3$.
We adopt as baseline the \textbf{CCA}~\cite{Rasiwasia:2010:NAC:1873951.1873987} method, a linear subspace learning approach, and non-linear neural-based methods, \textbf{Bi-AE}~\cite{Ngiam:2011:MDL:3104482.3104569}, \textbf{Bi-DBN}~\cite{Srivastava12learningrepresentations}, \textbf{Corr-AE}~\cite{Feng:2014:CRC:2647868.2654902}, \textbf{Corr-Cross-AE}~\cite{Feng:2014:CRC:2647868.2654902}, \textbf{Corr-Full-AE}~\cite{Feng:2014:CRC:2647868.2654902}, and \textbf{DCCA}~\cite{Andrew:2013:DCC:3042817.3043076,7298966}. All baselines are atemporal. We use the TempXNet with the three temporal smoothing functions.  For all datasets, we use $90\%$ of the data for development and the remaining for testing. We further split the development data using $15\%$ for validation. We consider \emph{days} as the temporal granularity for social stories and \emph{years} for NUS-WIDE.

All methods are evaluated on the three datasets, of varying dimensions, representing corpora with different topic broadness, and thus distinct temporal dynamics. Figure~\ref{table:retrieval_nuswide}, Figure~\ref{table:retrieval_edfest} and Figure~\ref{table:retrieval_tdf}, show the $mAP@50$ and $nDCG@50$ results for the NUS-WIDE, EdFest2016 and TDF2016 datasets, respectively.

The first observation we draw from the results is that TempXNet is highly effective across the three datasets, outperforming all the baselines, on both tasks, on all metrics. Specifically, TempXNet is able to rank at the top ($nDCG$) highly relevant instances (i.e. instances that share more semantic categories). This confirms our hypothesis regarding modelling temporal correlations, through temporal subspace learning. 

\begin{figure*}[t]
\minipage{0.21\linewidth}
  \includegraphics[width=\linewidth]{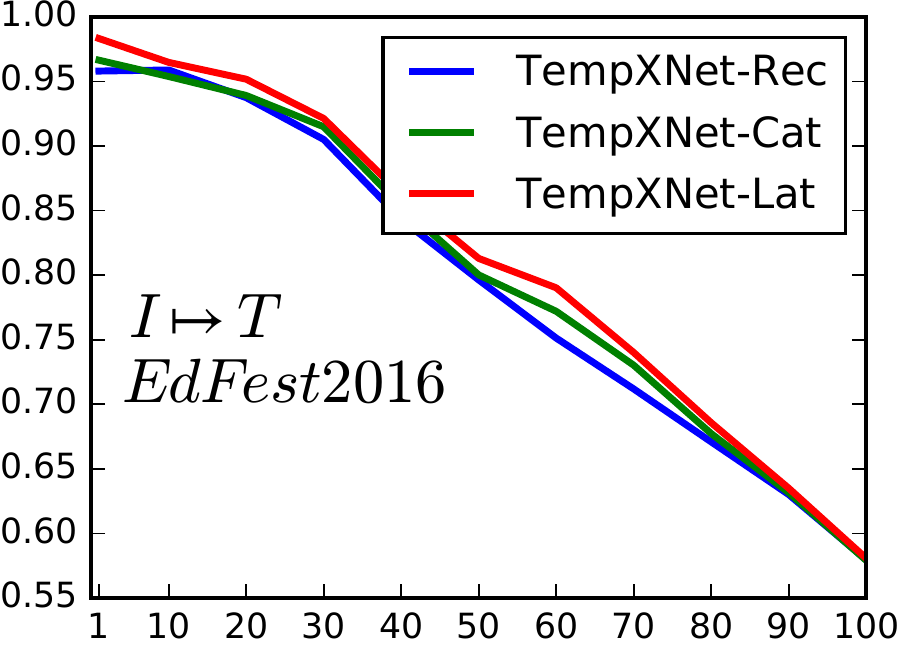}
\endminipage
\hspace{2mm}
\minipage{0.21\linewidth}
  \includegraphics[width=\linewidth]{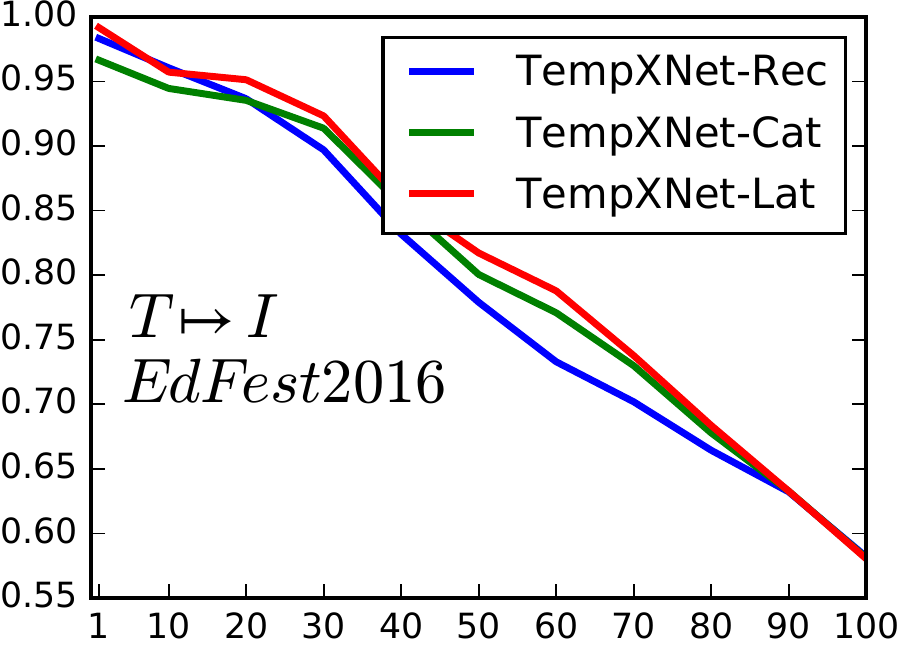}
\endminipage
\hspace{2mm}
\minipage{0.21\linewidth}
  \includegraphics[width=\linewidth]{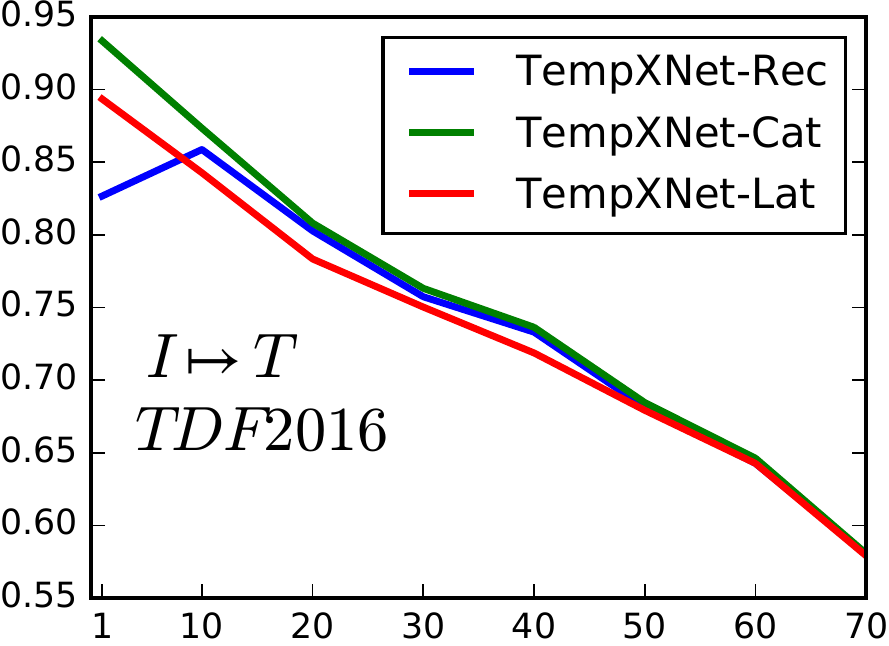}
\endminipage
\hspace{2mm}
\minipage{0.21\linewidth}
  \includegraphics[width=\linewidth]{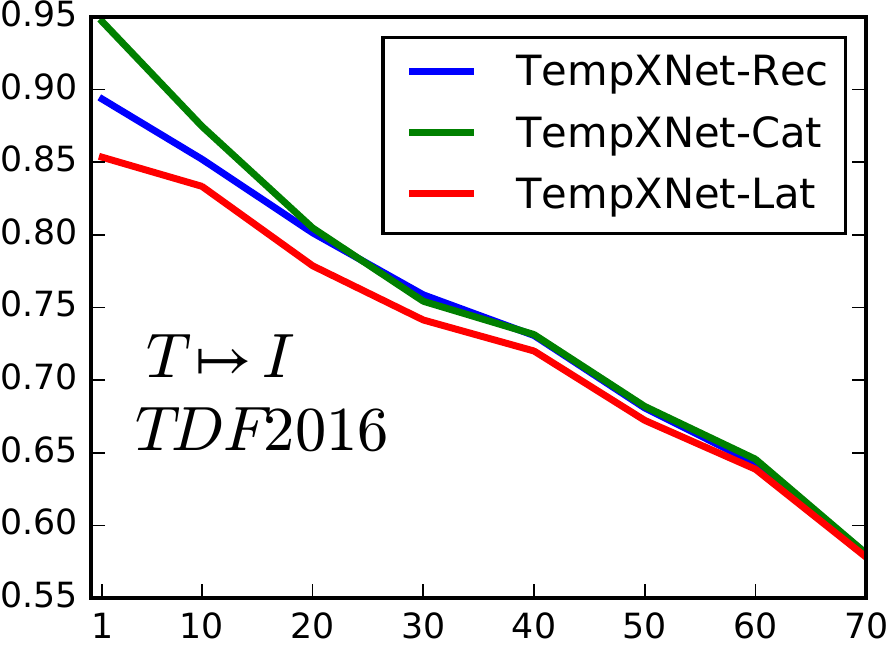}
\endminipage
\vspace{-8pt}
\caption{Precision-Scope curves for EdFest2016 and TDF2016.}
\label{fig:scope}
\end{figure*}

Regarding the different temporal smoothing functions, apart from NUS-WIDE, where distinct temporal correlations achieved identical performance, for EdFest2016 and TDF2016 performance oscillate. This reflects the existence of distinct temporal distributions, underlying each dataset. TempXNet-Lat outperforms the other correlations on EdFest2016. As TempXNet-Lat exploits temporal correlations of words, it is able to capture correlations between instances based on word's temporal behaviour. Additionally, on EdFest2016, TempXNet-Rec outperforms TempXNet-Cat. This indicates that for EdFest2016, latent-based and recency-based temporal correlations are more preferred, instead of category-based correlations. This means that: words temporal behaviour, for this particular dataset, helps discriminating instances, and instances that occur close to the query timestamp are more relevant. Such behaviour is expected when there are sporadic sub-events, provoking shifts on word's usage. 

On TDF2016 dataset, TempXNet-Cat outperforms all the other baselines and correlations by a considerable margin. This result indicates that for this dataset, focusing on semantic categories temporal density distributions helps retrieving more relevant content. This may be due to the existence of distributions with multiple modes (e.g. periodic dynamic behaviour). In fact, TDF2016 topics are to some extent periodic, e.g. stages, mountain races, news regarding winners, etc.

\begin{table}[t]
\caption{Comparison on Cross-Media retrieval ($mAP@50$ and $nDCG@50$) on NUS-WIDE.}
\vspace{-10pt}
\label{table:retrieval_nuswide}
\centering
\resizebox{\columnwidth}{!}{%
\begin{tabular}{|l|cccccc|}
\hline
\multirow{2}{*}{Method~~~~\%} & \multicolumn{2}{c|}{$I\mapsto  T$} &
\multicolumn{2}{c|}{$T\mapsto I$} &
\multicolumn{2}{c|}{Avg.}\\
\cline{2-7}
& \small{$mAP$} & \multicolumn{1}{c|}{\small{$nDCG$}} & \small{$mAP$} & \multicolumn{1}{c|}{\small{$nDCG$}}& \small{$mAP$} & \multicolumn{1}{c|}{\small{$nDCG$}}
\\
\hline\hline
{\small{CCA~\cite{Rasiwasia:2010:NAC:1873951.1873987}}} &
\multicolumn{1}{c}{74.2} &  
\multicolumn{1}{c|}{84.4} & 
\multicolumn{1}{c}{68.7} & 
\multicolumn{1}{c|}{80.7} &  
\multicolumn{1}{c}{71.5} & 
\multicolumn{1}{c|}{82.6}\\
\hline
{\small{Bi-AE~\cite{Ngiam:2011:MDL:3104482.3104569}}} & 
\multicolumn{1}{c}{74.1} &  
\multicolumn{1}{c|}{84.9} & 
\multicolumn{1}{c}{69.1} & 
\multicolumn{1}{c|}{80.0} &  
\multicolumn{1}{c}{71.6} & 
\multicolumn{1}{c|}{82.4}\\
\hline
{\small{Bi-DBN~\cite{Srivastava12learningrepresentations}}} & 
\multicolumn{1}{c}{69.5} &  
\multicolumn{1}{c|}{81.7} & 
\multicolumn{1}{c}{53.7} & 
\multicolumn{1}{c|}{67.8} &  
\multicolumn{1}{c}{61.6} & 
\multicolumn{1}{c|}{74.7}\\
\hline
{\small{Corr-AE~\cite{Feng:2014:CRC:2647868.2654902}}} & 
\multicolumn{1}{c}{76.2} &  
\multicolumn{1}{c|}{86.3} & 
\multicolumn{1}{c}{74.3} & 
\multicolumn{1}{c|}{83.9} &  
\multicolumn{1}{c}{75.2} & 
\multicolumn{1}{c|}{85.1}\\
\hline
{\small{Corr-Cross-AE~\cite{Feng:2014:CRC:2647868.2654902}}} & 
\multicolumn{1}{c}{72.8} &  
\multicolumn{1}{c|}{84.4} & 
\multicolumn{1}{c}{74.8} & 
\multicolumn{1}{c|}{84.4} &  
\multicolumn{1}{c}{73.8} & 
\multicolumn{1}{c|}{84.4}\\
\hline
{\small{Corr-Full-AE~\cite{Feng:2014:CRC:2647868.2654902}}} &
\multicolumn{1}{c}{75.4} &  
\multicolumn{1}{c|}{86.0} & 
\multicolumn{1}{c}{75.5} & 
\multicolumn{1}{c|}{84.6} &  
\multicolumn{1}{c}{75.5} & 
\multicolumn{1}{c|}{85.3}\\
\hline
{\small{DCCA~\cite{Andrew:2013:DCC:3042817.3043076,7298966}}} & 
\multicolumn{1}{c}{73.9} &  
\multicolumn{1}{c|}{85.1} & 
\multicolumn{1}{c}{76.1} & 
\multicolumn{1}{c|}{85.0} &  
\multicolumn{1}{c}{75.0} & 
\multicolumn{1}{c|}{85.1}\\
\hline\hline
\small{TempXNet-Rec} & 
\multicolumn{1}{c}{78.7} &  
\multicolumn{1}{c|}{86.6} & 
\multicolumn{1}{c}{79.9} & 
\multicolumn{1}{c|}{87.6} &  
\multicolumn{1}{c}{79.3} & 
\multicolumn{1}{c|}{87.1} \\
\hline
\small{TempXNet-Cat} & 
\multicolumn{1}{c}{78.8} &  
\multicolumn{1}{c|}{86.6} & 
\multicolumn{1}{c}{\textbf{80.0}} & 
\multicolumn{1}{c|}{\textbf{87.7}} &  
\multicolumn{1}{c}{\textbf{79.4}} & 
\multicolumn{1}{c|}{\textbf{87.2}} \\
\hline
\small{TempXNet-Lat} & 
\multicolumn{1}{c}{\textbf{79.1}} &  
\multicolumn{1}{c|}{\textbf{86.9}} & 
\multicolumn{1}{c}{79.5} & 
\multicolumn{1}{c|}{87.4} &  
\multicolumn{1}{c}{79.3} & 
\multicolumn{1}{c|}{\textbf{87.2}} \\
\hline
\end{tabular}%
}
\end{table}
\begin{table}[t]
\caption{Comparison on Cross-Media retrieval ($mAP@50$ and $nDCG@50$) on EdFest2016.}
\vspace{-10pt}
\label{table:retrieval_edfest}
\centering
\resizebox{\columnwidth}{!}{%
\begin{tabular}{|l|cccccc|}
\hline
\multirow{2}{*}{Method~~~~\%} & \multicolumn{2}{c|}{$I\mapsto T$} &
\multicolumn{2}{c|}{$T\mapsto I$} &
\multicolumn{2}{c|}{Avg.}\\
\cline{2-7}
& \small{$mAP$} & \multicolumn{1}{c|}{\small{$nDCG$}} & \small{$mAP$} & \multicolumn{1}{c|}{\small{$nDCG$}}& \small{$mAP$} & \multicolumn{1}{c|}{\small{$nDCG$}}
\\
\hline\hline
{\small{CCA~\cite{Rasiwasia:2010:NAC:1873951.1873987}}} & 
\multicolumn{1}{c}{58.6} &  
\multicolumn{1}{c|}{75.5} & 
\multicolumn{1}{c}{53.3} & 
\multicolumn{1}{c|}{73.7} &  
\multicolumn{1}{c}{56.0} & 
\multicolumn{1}{c|}{74.6} \\
\hline
{\small{Bi-AE~\cite{Ngiam:2011:MDL:3104482.3104569}}} & 
\multicolumn{1}{c}{64.9} &  
\multicolumn{1}{c|}{83.8} & 
\multicolumn{1}{c}{66.4} & 
\multicolumn{1}{c|}{83.0} &  
\multicolumn{1}{c}{65.7} & 
\multicolumn{1}{c|}{83.4}\\
\hline
{\small{Bi-DBN~\cite{Srivastava12learningrepresentations}}} & 
\multicolumn{1}{c}{56,7} &  
\multicolumn{1}{c|}{78.3} & 
\multicolumn{1}{c}{46.7} & 
\multicolumn{1}{c|}{67.1} &  
\multicolumn{1}{c}{51.7} & 
\multicolumn{1}{c|}{72.7}\\
\hline
{\small{Corr-AE~\cite{Feng:2014:CRC:2647868.2654902}}} & 
\multicolumn{1}{c}{67.8} &  
\multicolumn{1}{c|}{85.8} & 
\multicolumn{1}{c}{67.8} & 
\multicolumn{1}{c|}{83.0} &  
\multicolumn{1}{c}{67.8} & 
\multicolumn{1}{c|}{84.4}\\
\hline
{\small{Corr-Cross-AE~\cite{Feng:2014:CRC:2647868.2654902}}} & 
\multicolumn{1}{c}{60.0} &  
\multicolumn{1}{c|}{80.6} & 
\multicolumn{1}{c}{64.3} & 
\multicolumn{1}{c|}{81.4} &  
\multicolumn{1}{c}{62.2} & 
\multicolumn{1}{c|}{81.0}\\
\hline
{\small{Corr-Full-AE~\cite{Feng:2014:CRC:2647868.2654902}}} & 
\multicolumn{1}{c}{68.0} &  
\multicolumn{1}{c|}{85.4} & 
\multicolumn{1}{c}{68.7} & 
\multicolumn{1}{c|}{83.2} &  
\multicolumn{1}{c}{68.3} & 
\multicolumn{1}{c|}{84.3}\\
\hline
{\small{DCCA~\cite{Andrew:2013:DCC:3042817.3043076,7298966}}} & 
\multicolumn{1}{c}{89.7} &  
\multicolumn{1}{c|}{96.2} & 
\multicolumn{1}{c}{72.4} & 
\multicolumn{1}{c|}{85.5} &  
\multicolumn{1}{c}{81.1} & 
\multicolumn{1}{c|}{90.9}\\
\hline\hline
\small{TempXNet-Rec} & 
\multicolumn{1}{c}{94.5} &  
\multicolumn{1}{c|}{97.4} & 
\multicolumn{1}{c}{\textbf{95.5}} & 
\multicolumn{1}{c|}{97.7} &  
\multicolumn{1}{c}{95.0} & 
\multicolumn{1}{c|}{97.6}\\
\hline
\small{TempXNet-Cat} & 
\multicolumn{1}{c}{94.0} &  
\multicolumn{1}{c|}{96.9} & 
\multicolumn{1}{c}{93.6} & 
\multicolumn{1}{c|}{97.3} &  
\multicolumn{1}{c}{93.8} & 
\multicolumn{1}{c|}{97.1} \\
\hline
\small{TempXNet-Lat} & 
\multicolumn{1}{c}{\textbf{96.4}} &  
\multicolumn{1}{c|}{\textbf{98.6}} & 
\multicolumn{1}{c}{\textbf{95.5}} & 
\multicolumn{1}{c|}{\textbf{98.1}} &  
\multicolumn{1}{c}{\textbf{96.0}} & 
\multicolumn{1}{c|}{\textbf{98.4}}\\
\hline
\end{tabular}%
}
\end{table}

\begin{table}[t]
\caption{Comparison on Cross-Media retrieval ($mAP@50$ and $nDCG@50$) on TDF2016.}
\vspace{-10pt}
\label{table:retrieval_tdf}
\centering
\resizebox{\columnwidth}{!}{%
\begin{tabular}{|l|cccccc|}
\hline
\multirow{3}{*}{Method~~~~\%}   & \multicolumn{2}{c|}{$I\mapsto T$} &
\multicolumn{2}{c|}{$T\mapsto I$} &
\multicolumn{2}{c|}{Avg.}\\
\cline{2-7}
& \small{$mAP$} & \multicolumn{1}{c|}{\small{$nDCG$}} & \small{$mAP$} & \multicolumn{1}{c|}{\small{$nDCG$}}& \small{$mAP$} & \multicolumn{1}{c|}{\small{$nDCG$}}
\\
\hline\hline
{\small{CCA~\cite{Rasiwasia:2010:NAC:1873951.1873987}}} & 
\multicolumn{1}{c}{58.0} &  
\multicolumn{1}{c|}{76.9} & 
\multicolumn{1}{c}{57.7} & 
\multicolumn{1}{c|}{75.4} &  
\multicolumn{1}{c}{57.8} & 
\multicolumn{1}{c|}{76.2}\\
\hline
{\small{Bi-AE~\cite{Ngiam:2011:MDL:3104482.3104569}}} & 
\multicolumn{1}{c}{72.5} & 
\multicolumn{1}{c|}{88.6} & 
\multicolumn{1}{c}{67.0} & 
\multicolumn{1}{c|}{82.2} &   
\multicolumn{1}{c}{69.7} & 
\multicolumn{1}{c|}{85.5}\\
\hline
{\small{Bi-DBN~\cite{Srivastava12learningrepresentations}}} & 
\multicolumn{1}{c}{64.5} &  
\multicolumn{1}{c|}{82.9} & 
\multicolumn{1}{c}{56.1} & 
\multicolumn{1}{c|}{74.2} &  
\multicolumn{1}{c}{60.3} & 
\multicolumn{1}{c|}{78.6}\\
\hline
{\small{Corr-AE~\cite{Feng:2014:CRC:2647868.2654902}}} & 
\multicolumn{1}{c}{73.5} &  
\multicolumn{1}{c|}{89.1} & 
\multicolumn{1}{c}{71.4} & 
\multicolumn{1}{c|}{86.1} &  
\multicolumn{1}{c}{72.4} & 
\multicolumn{1}{c|}{87.6}\\
\hline
{\small{Corr-Cross-AE~\cite{Feng:2014:CRC:2647868.2654902}}} & 
\multicolumn{1}{c}{70.5} &  
\multicolumn{1}{c|}{85.9} & 
\multicolumn{1}{c}{72.2} & 
\multicolumn{1}{c|}{86.3} &  
\multicolumn{1}{c}{71.4} & 
\multicolumn{1}{c|}{86.0}\\
\hline
{\small{Corr-Full-AE~\cite{Feng:2014:CRC:2647868.2654902}}} & 
\multicolumn{1}{c}{74.1} &  
\multicolumn{1}{c|}{89.4} & 
\multicolumn{1}{c}{71.8} & 
\multicolumn{1}{c|}{86.5} &  
\multicolumn{1}{c}{73.0} & 
\multicolumn{1}{c|}{88.0}\\
\hline
{\small{DCCA~\cite{Andrew:2013:DCC:3042817.3043076,7298966}}} & 
\multicolumn{1}{c}{88.4} &  
\multicolumn{1}{c|}{95.5} & 
\multicolumn{1}{c}{73.8} & 
\multicolumn{1}{c|}{86.2} &  
\multicolumn{1}{c}{81.1} & 
\multicolumn{1}{c|}{90.9}\\
\hline\hline
\small{TempXNet-Rec} & 
\multicolumn{1}{c}{87.2} &  
\multicolumn{1}{c|}{93.9} & 
\multicolumn{1}{c}{89.1} & 
\multicolumn{1}{c|}{94.6} &  
\multicolumn{1}{c}{88.2} & 
\multicolumn{1}{c|}{94.3}\\
\hline
\small{TempXNet-Cat} & 
\multicolumn{1}{c}{\textbf{92.6}} &  
\multicolumn{1}{c|}{\textbf{96.8}} & 
\multicolumn{1}{c}{\textbf{91.5}} & 
\multicolumn{1}{c|}{\textbf{95.9}} &  
\multicolumn{1}{c}{\textbf{92.1}} & 
\multicolumn{1}{c|}{\textbf{96.4}}\\
\hline
\small{TempXNet-Lat} & 
\multicolumn{1}{c}{88.1} &  
\multicolumn{1}{c|}{94.7} & 
\multicolumn{1}{c}{90.3} & 
\multicolumn{1}{c|}{95.8} &  
\multicolumn{1}{c}{89.2} & 
\multicolumn{1}{c|}{95.3}\\
\hline
\end{tabular}%
}
\end{table}

Figure~\ref{fig:scope} shows the precision-scope curves for both EdFest2016 and TDF2016 datasets, on the Image-to-Text and Text-to-Image tasks. On the $x$ axis we vary the value of $k$, and the $y$ axis shows the corresponding $mAP@k$. On EdFest2016, it can be observed that TempXNet-Lat always outperforms the remaining correlations. Similarly, on TDF2016 TempXNet-Cat also outperforms the remaining correlations, which is consistent with the previously discussed results. 

In the presence of datasets with different intrinsic temporal distributions, our temporal cross-media subspace learning model is able to effectively model such distributions, provided that a suitable temporal correlation is used. Apart from the three  temporal correlations evaluated, TempXNet can accommodate any other temporal distributions, as long as they can be expressed through a differentiable function.

\begin{figure*}[t]

\includegraphics[width=0.70\linewidth]{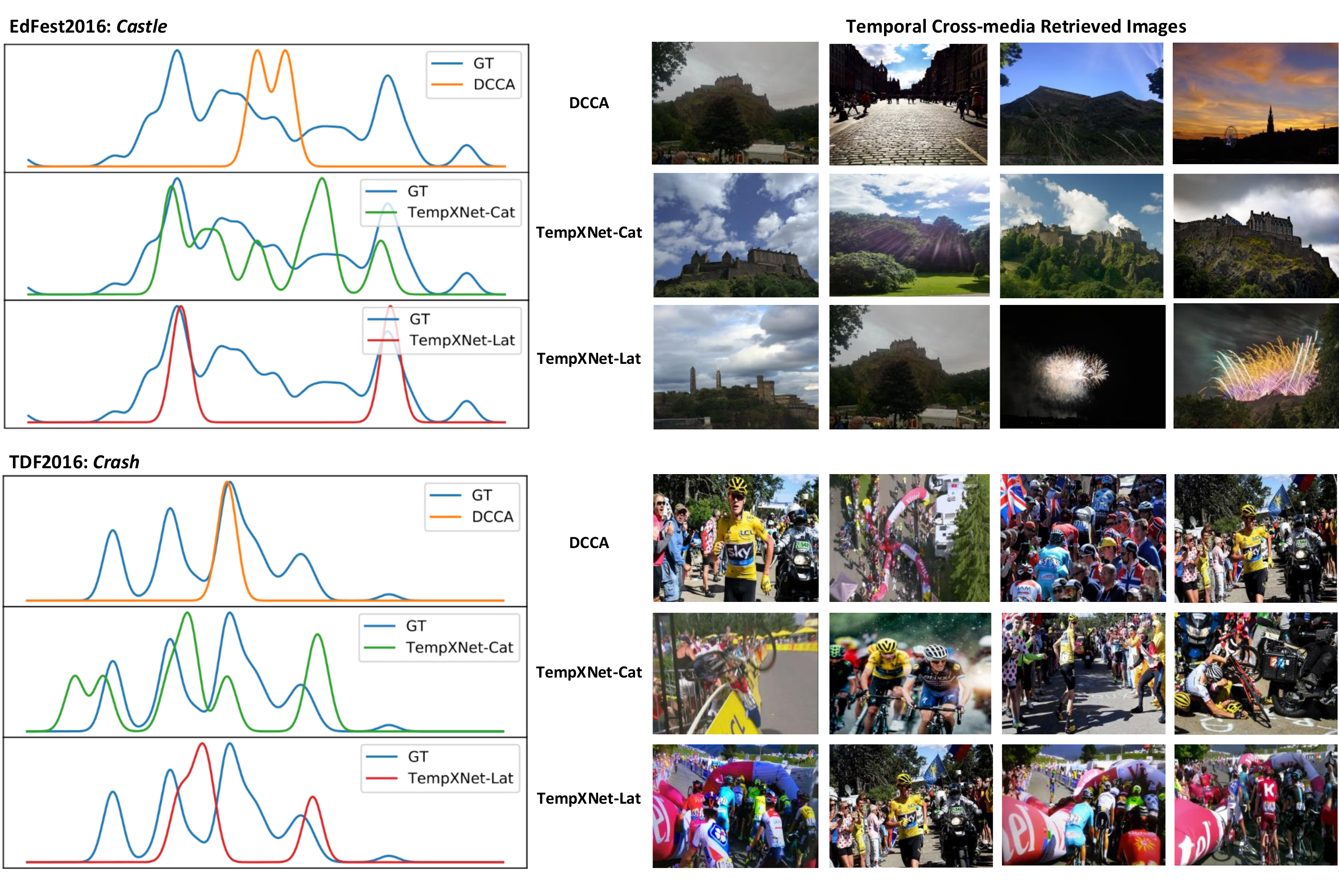}
\vspace{-4mm}
\caption{Quantitative analysis of the different temporal correlations on the EdFest2016 and TDF2016 dataset. Each plot depicts the temporal distribution of ground-truth instances, from the categories \emph{Castle} and \emph{Crash}. We use \emph{days} as time granularity.}
\label{fig:qualitative}
\end{figure*}

\vspace{5mm}
\textbf{Media temporal correlations.}
In this section we perform a qualitative analysis of the different temporal correlations. The goal is to assess how well temporal correlations are captured by each temporal model. To this end, we query each model and compare its relevant instances distribution with the true ground-truth temporal distribution. Specifically, we perform two queries, one for EdFest2016 in which the target are instances of the semantic category  \emph{Castle} and one for TDF2016 in which the target are instances of \emph{Crash}, respectively. Each query comprises only the textual modality, corresponding to the $T\mapsto I$ setting. The two top performing temporal correlations (TempXNet-Cat and TempXNet-Lat) and the DCCA baseline are considered. For each query we show four sample images. Figure~\ref{fig:qualitative} depicts the result of this experiment. Each plot depicts the temporal distribution of ground-truth (GT) and relevant instances retrieved by each model, with the $x$-axis corresponding to time. We normalise each distribution to the $[0,1]$ range.

On the EdFest2016 plot, one can observe that the temporal distribution of the semantic category \emph{Castle} has multiple peaks, with the larger ones being on the borders. These correspond to the begginning and ending of the festival, in which at the end a fireworks show takes place. Although the temporal correlations are different, it can be seen that both TempXNet-Cat and TempXNet-Lat are able to cover both larger peaks, by retrieving relevant instances at the corresponding moments in time. Even though TempXNet-Cat achieved a better fit to GT, TempXNet-Lat achieved better retrieval results. This is may be due to the fact it covers the most salient peaks. On TDF2016 there were several crashes during the event, and this is reflected in the peaks of the ground-truth. Given the somewhat periodic nature of these peaks, TempXNet-Cat reveals a better fit to the GT curve. The fit of the TempXNet-Lat correlation is slightly worse, as it is based on individual word dynamics, and despite the periodic shape of the category \emph{Crash}, words that occur in \emph{Crash} instances may not reveal this pattern (e.g. usually it refers to racers names, etc.). The DCCA baseline completely fails to capture the temporal distribution of relevant documents. Given this observations, we verify that our model can effectively grasp temporal correlations of data.

\section{Conclusions}
In this paper we looked into the important problem of modelling semantically similar media that vary over time.
Current state-of-the-art cross-media methods assume that collections are static, overlooking visual and textual correlations (and cross-correlations) that change over time. TempXNet was thoroughly evaluated, exposing four fundamental concluding points.

\textbf{Temporal cross-media subspace.} A novel approach to cross-media retrieval was successfully proposed. It derives from the idea that multimedia data should be organized according to their semantic category and temporal correlations across different modalities. Several key components make the creation of this subspace possible.

\textbf{Principled temporal soft-constraints.} The creation of the subspace is temporally constrained by estimating temporal correlations of semantic categories and words, encoding the underlying dynamics of modalities. 
The investigated forms of soft-constraints stem from well-grounded statistical principles leading to a solid and rigorous optimisation framework. Hence, modality projections are learned through two coupled neural networks that are jointly optimised, subject to the aforementioned temporal constraints.

\textbf{Models of temporal cross-media correlations.} We observed that temporal correlations are seldomly simple as the recency model of temporal correlations was never the best model. In fact, we could contrast the results in the EdFest2016 and the TDF2016 datasets and conclude that both datasets follow different distributions: the EdFest2016 has several one time shows and events, and the TDF2016 contains several repeated events.

\textbf{Improved retrieval precision in dynamic domains.} Accounting for temporal cross-media correlations improved cross-modality retrieval across all datasets. The proposed TempXNet models outperformed past cross-media models. Moreover, the best retrieval precision was obtained by the TempXNet-Cat and TempXNet-Lat that model temporal correlations with different levels of granularity.

\begin{acks}
This work has been partially funded by the \grantsponsor{EU H2020}{H2020 ICT}{} project COGNITUS with the grant agreement number \grantnum{EU H2020}{687605} and by the \grantsponsor{FCT}{FCT}{} project NOVA LINCS Ref. \grantnum{FCT}{UID/CEC/04516/2013}. We also gratefully acknowledge the support of NVIDIA Corporation with the donation of the GPUs used for this research.
\end{acks}

\clearpage
\balance
\bibliographystyle{ACM-Reference-Format}

\end{document}